\pdfoutput=1
\documentclass[aps,pre,twocolumn,nofootinbib,superscriptaddress]{revtex4}
\usepackage[english]{babel}
\usepackage[utf8]{inputenc}
\usepackage{amsmath}
\usepackage{amssymb}
\usepackage{url}

\usepackage{graphicx}
\usepackage{float}
\usepackage{multirow}

\begin{document}

\author{J. Varennes}
\affiliation{Department of Physics and Astronomy, Purdue University, West Lafayette, IN 47907, USA}

\author{B. Han}
\affiliation{Schools of Mechanical Engineering \& Biomedical Engineering, Purdue University, West Lafayette, IN 47907, USA}

\author{A. Mugler}
\email{amugler@purdue.edu}
\affiliation{Department of Physics and Astronomy, Purdue University, West Lafayette, IN 47907, USA}

\title{Collective chemotaxis through noisy multicellular gradient sensing}

\begin{abstract}
Collective cell migration in response to a chemical cue occurs in many biological processes such as morphogenesis and cancer metastasis. Clusters of migratory cells in these systems are capable of responding to gradients of less than 1\% difference in chemical concentration across a cell length. Multicellular systems are extremely sensitive to their environment and while the limits to multicellular sensing are becoming known, how this information leads to coherent migration remains poorly understood. We develop a computational model of multicellular sensing and migration in which groups of cells collectively measure noisy chemical gradients. The output of the sensing process is coupled to individual cells’ polarization to model migratory behavior. Through the use of numerical simulations, we find that larger clusters of cells detect the gradient direction with higher precision and thus achieve stronger polarization bias, but larger clusters also induce more drag on collective motion. The trade-off between these two effects leads to an optimal cluster size for most efficient migration. We discuss how our model could be validated using simple, phenomenological experiments.
\end{abstract}

\maketitle

\newpage


\section{Introduction}

Cells can migrate in response to a chemoattractant and can detect extraordinarily small changes in chemical concentrations. The limits to cell sensory precision have been a topic of research in biology and biophysics for many years. \textit{Escheria coli} bacterial chemotaxis operates very near the physical limits of their sensory machinery, and \textit{Dictyostelium discoideum} amoebae are sensitive to differences in chemical concentrations on the order of ten molecules across the cell \cite{berg1977physics,song2006dictyostelium}. Recent studies on individual breast cancer cells showed that they are sensitive to 1\% differences in concentration across the cell length \cite{shields2007autologous}. Limits to cell sensory precision were first derived by Berg and Purcell almost 40 years ago \cite{berg1977physics} and have been revisited to account for binding kinetics, spatiotemporal correlations and spatial confinement \cite{bialek2005physical, kaizu2014berg, bicknell2015limits}. However, in nature cells are rarely found alone, and the interactions between nearby cells may alter cells' sensory capabilities.

In many biological contexts cells act in close proximity to one another which can have significant effects on collective behavior. Clusters of mammary epithelial cells, lymphocytes and neural crest cells can detect chemical gradients that single cells cannot \cite{ellison2016cell,malet2015collective,theveneau2010collective}, and cultures of neurons have been shown to be sensitive to single molecule differences across an individual neuron's axonal growth cone \cite{rosoff2004new}. In many types of cancer, tumor cell invasion is collective, involving coherent grouped motion guided by chemical cues \cite{cheung2013collective, friedl2012classifying, aceto2014circulating, puliafito2015three}. It is clear from these examples that cells acting collectively can improve upon their individual sensory precision. Similar to the limits set by Berg and Purcell, the physical limits to collective gradient sensing have been recently derived \cite{mugler2016limits,ellison2016cell} by using a multicellular version of the local excitation-global inhibition (LEGI) communication model \cite{levchenko2002models}, one of the simplest adaptive mechanisms of gradient sensing. With these studies the physical limits of cell sensing have been extended from single cells to multicellular collectives.

In parallel to research on cell sensory precision, studies on collective cell  migration have also advanced. Biological processes such as development, cellular migration, pathogenic response, and cancer progression all involve many cells acting in a coordinated way \cite{scarpa2016collective,friedl2010plasticity,rasmussen2006quorum,boelens2014exosome,cheung2013collective,vader2014extracellular,szabo2016modelling}. Simple mechanical models successfully explain observed collective behaviors such as cell streaming, cell sorting, cell sheet migration, wound healing, and cell aggregation \cite{kabla2012collective,szabo2010collective,basan2013alignment,janulevicius2015short}. These models accurately model collective cell migration but fail to explicitly include the affects of multicellular sensing in driving the mechanics at play. Cells are often capable of intercellular communication, so understanding how communicated information is translated into mechanical action is of prime interest.

How the phenomena of collective sensing and multicellular migration are connected remains an open question \cite{varennes2016sense,defranco2008migrating,haeger2015collective}. Recent studies by Camley et al.\ \cite{camley2016emergent} and Malet-Engra et al.\ \cite{malet2015collective} have started to address this need for modeling collective sensing and migration. In the study of Camley et al.\ individual cell measurements act to polarize cells in a cluster outwards causing tension, and when intercellular communication is added the tension on the cluster adapts to the chemical concentration. Both studies do not take into account the inherent stochasticity of cell sensing and intercellular communication. However, individual cell measurements of the environment are error-prone while propagation of single cell measurements also adds noise to the system. These studies also treat cells or clusters as perfect circles, neglecting natural geometric fluctuations in the size and shapes of cells that occur during migration.

Here we focus our attention on stochastic processes governing collective gradient sensing and cell motility. First, the limits to collective gradient sensing are briefly reviewed and our implementation of multicellular LEGI described. Information gained from collective sensing then must be used to direct cell motion. We develop a model which takes into account the fluctuating shape of cells while coupling cell motility to noisy collective gradient sensing. We model intercellular communication via the direct exchange of messenger molecules between cells. Candidate mediators of such intercellular communication have been recently identified in \textit{Drosophila} development \cite{ramel2013rab11}, and other studies suggest intercellular communication's involvement in organoid branching, angiogenesis, and cancer \cite{ellison2016cell,gerhardt2003vegf,hsu2000cadherin,friedl2009collective}. We study cluster migration in shallow gradients where the change in concentration across a cell width is very small relative to the background concentration. This regime is of prime interest since experiments show that collectives can respond to these shallow gradients whereas single cells cannot \cite{ellison2016cell,malet2015collective,rosoff2004new}. By explicitly modeling the stochastic processes of sensing and migration this model places constraints on the collective behavior of cells and predicts an optimal cluster size for fastest chemotaxis. We conclude by discussing our model's implications for cell migration experiments.

\begin{figure*}
    \centering
        \includegraphics[width=.75\textwidth]{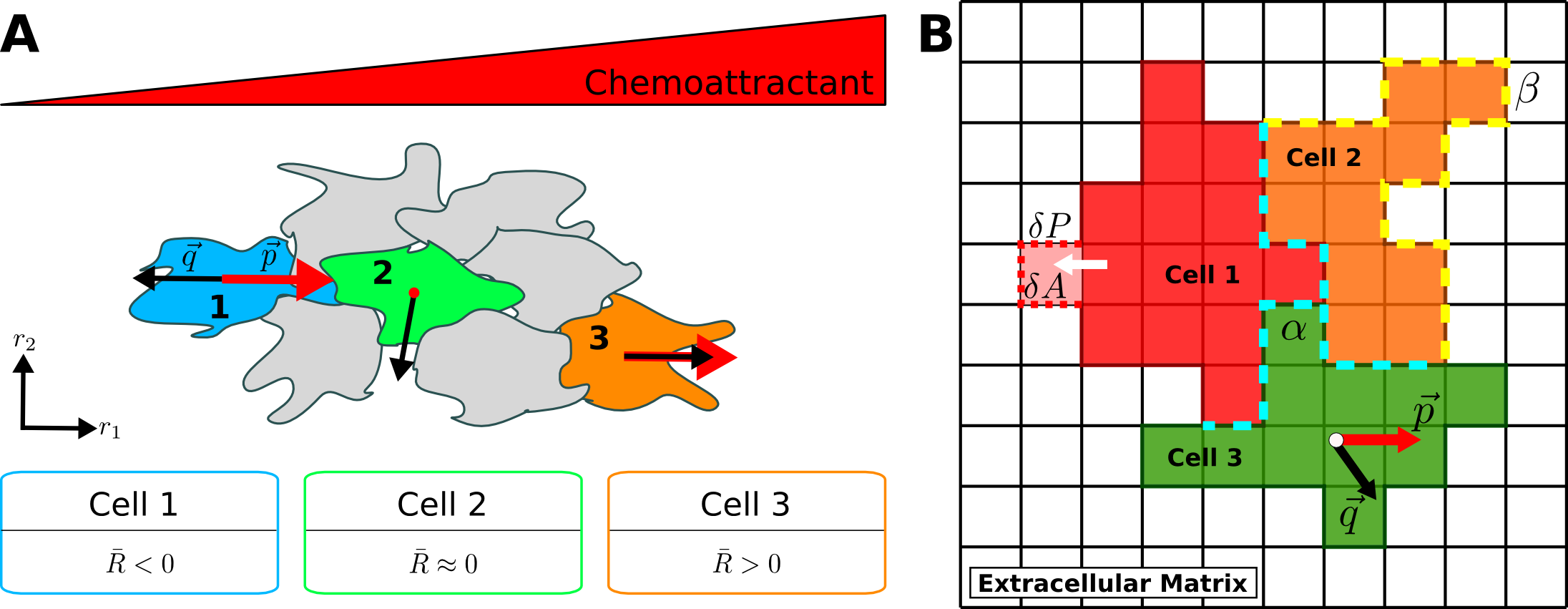}
    \caption{Model implementation. (A) Cell polarization is biased by multicellular sensing. On average, the cells on the left and right edges will measure negative and positive values of $R$, respectively. This causes the left-edge (Cell 1) and right-edge (Cell 3) cells to polarize in the direction of the gradient, while cells in the middle (Cell 2) are on average not polarized since $\bar{R} \approx 0$. Polarization vectors $\vec{p}$ are red, repulsion vectors $\vec{q}$ are black. (B) Simulations are implemented using the Cellular Potts Model (CPM). Cells comprise of simply connected lattice points. There are adhesion energies associated with different types of contact: cell-cell, $\alpha$ (blue-dashed line), and cell-ECM, $\beta$ (yellow-dashed line). Cell motility is modeled by the addition/removal of lattice points (pink). Each cell has a center-of-mass (white dot), a polarization vector, $\vec{p}$ (red) and a repulsion vector, $\vec{q}$ (black).} \label{fig:model}
\end{figure*}

\section{MODEL}

Communication between cells and collective sensing can improve upon an individual cell's ability to sense the environment \cite{ellison2016cell}, and in turn this information may be used to direct cell motion. To describe collective sensing, we will use the well-established local excitation--global inhibition (LEGI) mechanism \cite{mugler2016limits,levchenko2002models}.

\subsection{Limits to Multicellular Sensing}

Individual cells measure spatial gradients by comparing concentration measurements $c$ made by receptors or groups of receptors on the cell surface \cite{mugler2016limits,jilkine2011comparison}. For simplicity, we assume that a cell of size $a$ compares the number of diffusing molecules within two different regions of size $b$ which run parallel to the chemical gradient $\bar{g}$.
The relative error in each compartment's measurement is
$ \left(\sigma_c / \bar{c} \right)^2 \sim 1 / \left( b\bar{c}DT \right) $ \cite{berg1977physics},
where $D$ is the diffusion coefficient, and $T$ is the measurement integration time. Assuming that the measurements made in each compartment are independent, then the difference in counts is proportional to the gradient
$\Delta\bar{n} = \bar{n}_2 - \bar{n}_1 \sim ab^3\bar{g}$.
In the limit that the gradient is very small relative to the background concentration $a\bar{g}\ll\bar{c}$, the relative error in gradient sensing simplifies to
\begin{equation} \label{eq:g}
\frac{\sigma_g}{\bar{g}} = \frac{\sigma_{\Delta n}}{\Delta \bar{n}} \sim \sqrt{\frac{\bar{c}}{b(a\bar{g})^2DT}}.
\end{equation}
Eq.\ \ref{eq:g} has been extensively derived and generalized to systems with different geometries \cite{endres2008accuracy,endres2009accuracy,hu2010physical} and in all such cases a term of the form in Eq.\ \ref{eq:g} appears as the fundamental limit, with the length scale $b$ dictated by the particular sensory mechanism and geometry. In the case of multicellular gradient sensing, we consider the cells on opposite ends of a chain of cells as the two compartments comparing concentration measurements. Then in Eq.\ \ref{eq:g} $b \to a$ and $a \to Na$ where $N$ is the number of cells in the chain. The relative error for the multicellular cluster becomes \cite{mugler2016limits}
\begin{equation} \label{eq:G1}
\frac{\sigma_g}{\bar{g}} \sim \sqrt{\frac{\bar{c}}{a(Na\bar{g})^2DT}}.
\end{equation}
There is a crucial effect that is neglected in formulating Eq.\ \ref{eq:G1} which is the mechanism by which the cells communicate their measurements across the collective. Communication will introduce additional noise to the gradient sensing process thereby altering the expression for the relative error. In the case of a single cell it is reasonable to assume that measurements from different compartments can be reliably transmitted, but with the increased size of the multicellular cluster we cannot make the same assumption. Using the multicellular LEGI paradigm \cite{levchenko2002models} to model intercellular communication, the physical limits to communication-aided collective gradient sensing have been derived \cite{ellison2016cell, mugler2016limits}, which we expand upon below.

\subsection{Multicellular LEGI Model}

In the LEGI model cells produce two chemical species, a ``local'' species $X$, and a ``global'' species $Y$, in response to the chemoattractant $S$. The local species $X$ remains within an individual cell and represents that cell's measurement of its local chemical concentration. This species can be a molecule produced or activated in response to attractant-bound receptors, or the bound receptors themselves. The global species $Y$ can diffuse at the rate $\gamma$ between neighboring cells and therefore represents the average $X$ population among neighboring cells. $Y$ molecules may only be exchanged when two or more cells are in direct contact with one another. Recent experiments in epithelial cells identified this global species as either calcium or a small molecule involved in calcium signaling (such as IP3), and identified the intercell diffusion mechanism as mediated by gap junctions \cite{ellison2016cell}. Finally, $X$ activates a downstream reporter molecule $R$, while $Y$ inhibits $R$.

Let $x_k$, $y_k$, and $R_k$ represent the molecule populations in $X$, $Y$, and $R$ in the $k^\text{th}$ cell. The chemical reactions in cell $k$ are
\begin{equation}
    \begin{aligned}
        s_k &\xrightarrow{\kappa} s_k + x_k \hspace{20pt} x_k \xrightarrow{\mu} \emptyset \\
        s_k &\xrightarrow{\kappa} s_k + y_k \hspace{20pt} y_k \xrightarrow{\mu} \emptyset \hspace{20pt} y_k \rightleftharpoons_{\gamma_{j,k}}^{\gamma_{k,j}} y_{j} .
    \end{aligned}
\end{equation}
The production and degradation rates for $X$ and $Y$ are $\kappa$ and $\mu$, respectively. The global reporter molecule exchange rate $\gamma$ is dependent on the length of the interface $\mathcal{C}$ made between adjacent cells, and on the exchange rate per unit contact-length $\Gamma$,
$$\gamma_{j,k} = \int_{\mathcal{C}} \Gamma dl . $$
In the limit of strong communication ($\gamma \gg \mu$) and many cells, the relative error of gradient sensing is limited from below by \cite{mugler2016limits}
\begin{equation} \label{eq:G2}
\frac{\sigma_g}{\bar{g}} \sim \sqrt{\frac{\bar{c}}{a(n_0a\bar{g})^2DT}},
\end{equation}
where $n_0$ sets an effective number of cells over which information can be reliably conveyed. In our model communication between cells improves with increased diffusion of $Y$ molecules and so $n_0^2 \propto \gamma/\mu$ \cite{ellison2016cell, mugler2016limits}. As collectives grow larger than $n_0$ cells the relative error ceases to improve, saturating to the limit set by Eq.\ \ref{eq:G2}; unlike Eq.\ \ref{eq:G1} where the effects of communication are ignored and the relative error decreases without bound.

In the limit of shallow gradients, which are of primary interest in studying collective sensing, $R$ effectively reports the difference in $X$ and $Y$ molecule populations \cite{ellison2016cell} and so we will model the downstream readout as $R_k = x_k-y_k$. A negative (positive) difference indicates that the cell is below (above) the average measured concentration relative to nearby cells as shown by the reported average $R$ values for each cell in Fig.\ \ref{fig:model}A.

The chemical concentration is modeled as a space-dependent field $E(r_1,r_2)$, and in this case has a constant gradient in the $r_1$-direction,
\begin{gather*}
    E(r_1,r_2) = \bar{g}r_1 + \bar{c}.
\end{gather*}
The average signal in the $k^\text{th}$ cell's local environment is $\bar{s}_k = \int_{A_k} dr_1 dr_2 \ E(r_1,r_2)$ where $A_k$ is the area of the $k^\text{th}$ cell. Since diffusion is a Poisson process the variance in the measured signal $s_k$ is equal to the mean, $\sigma_{s_k}^2 = \bar{s}_k$. At each time step we sample $s_k$ for each cell from a Gaussian distribution with mean and variance $\bar{s}_k$, which corresponds to instantaneous sensory readout \cite{ellison2016cell}. The dynamics of the local reporter satisfy the stochastic differential equation
\begin{equation} \label{eq:xdot}
    \dot{x}_k = \kappa s_k - \mu x_k + \eta_{x_k}.
\end{equation}
The first term in Eq.\ \ref{eq:xdot} is due to the production of $X$ molecules due to the signal $S$, the second term represents molecule degradation, and the third term $\eta_{x_k}$ accounts for the noise inherent to these reactions. The noise term is equal to
$\eta_{x_k} = \sqrt{\kappa\bar{s}_k}\xi_{1,k} - \sqrt{\mu \bar{x}_k} \xi_{2,k}$ since both production and degradation are stochastic processes \cite{gillespie2000chemical}.
In Eq.\ \ref{eq:xdot} and subsequent stochastic equations $\xi_{i,k}$ and $\chi_{j,k}$ are unit Gaussian random variables representing the noise in molecule populations. For the local reporter, the steady-state solution is simply
\begin{equation} \label{eq:xss}
    x_{k}^{ss} = \left( \kappa/\mu \right) s_k + \left( 1/\mu \right) \eta_{x_k}.
\end{equation}

The dynamics of the global species can be modeled in similar fashion,
\begin{equation} \label{eq:ydot}
    \dot{y}_k = \kappa s_k - \mu y_k - y_k \sum_{\langle j,k \rangle} \gamma_{j,k} + \sum_{\langle j,k \rangle} y_j \ \gamma_{j,k} + \eta_{y_k} .
\end{equation}
The first summation term in Eq.\ \ref{eq:ydot} accounts for the loss of $y_k$ due to the diffusion out to neighboring cells, and similarly the second summation term accounts for the increase in $y_k$ due to diffusion into cell $k$ from its neighbors. The notation $\langle j,k \rangle$ represents the set of all nearest neighbor pairs. The noise term $\eta_{y_k}$ in the molecule dynamics depends on the production, degradation and diffusion of $Y$ molecules. In steady-state we can express the noise as
\begin{equation*}
    \eta_{y_k} = \sqrt{\kappa \bar{s}_k} \xi_4 - \sqrt{\mu\bar{y}_k} \xi_5 + \sum_{j=1}^N \left[ \chi_{j,k} \sqrt{\gamma_{j,k}} \left( \sqrt{\bar{y}_j}-\sqrt{\bar{y}_k} \right) \right] .
\end{equation*}
Similarly to $\eta_{x_k}$, the noise in $y_k$ also depends on production and degradation while an extra term is required to account for the noise in $Y$ molecule exchange. Eq.\ \ref{eq:ydot} can be simplified by noting that exchange rates between cells are symmetric $\gamma_{j,k}=\gamma_{k,j}$, $\gamma_{i,i}=0$, and by defining the sum of all the exchange rates between cell $k$ and all other cells as $G_k = \sum_{j=1}^N \gamma_{j,k}$. The steady-state solution for the global reporter is more involved than the local reporter, and can be written as a matrix equation
\begin{equation} \label{eq:yss}
    M\vec{y}^{ss} = \kappa\vec{s} + \vec{\eta}_y,
\end{equation}
where $M$ is a square, symmetric matrix that governs the degradation and exchange of $Y$ molecules in all cells,
\begin{equation} \label{eq:Mmatrix}
    M =
    \begin{bmatrix}
     \mu+G_1 & -\gamma_{1,2} & \cdots & -\gamma_{1,N} \\
     -\gamma_{2,1} & \mu+G_2 & \cdots & -\gamma_{2,N} \\
     \vdots  & \vdots  & \ddots & \vdots  \\
     -\gamma_{N,1} & -\gamma_{N,2} & \cdots & \mu+G_N
    \end{bmatrix}.
\end{equation}

\begin{table*}[b]
\centering
\footnotesize
\begin{tabular}{ |c|c|c| }
\hline
Parameter & Value & Notes \\ \hline
Concentration $\bar{c}$ & $10 \text{nM}$ & Assumes $\bar{c} \gg a\bar{g}$ for shallow gradients \cite{malet2015collective,ellison2016cell} \\
Gradient $\bar{g}$ & $0.04 \text{nM/}\mu\text{m}$ & \\ \hline
LEGI Molecule Production Rate $\kappa$ & $19.72 \text{min}^{-1}$ & Assumes $ \{\kappa,\mu\} \gg r$ \\
LEGI Molecule Degradation Rate $\mu$ & $19.72 \text{min}^{-1}$ & i.e.\ biochemical signaling is faster than motility response \\ \hline
Global Reporter Exchange Rate $\Gamma$ & $80 (\mu\text{m} \ \text{min})^{-1}$ & Varied in Fig.\ \ref{fig:results} \\ \hline
Polarization Bias Strength $\epsilon$ & 0.8 & Varied in Fig.\ \ref{fig:heat} \\ \hline
Polarization Decay Rate $r$ & $3.94 \text{min}^{-1}$ & Sets polarization memory time, as used in \cite{szabo2010collective} \\ \hline
Relaxed Cell Area $A_0$ & $315 \mu\text{m}^2$ & Assumes cell radius $10 \mu\text{m}$ \cite{leber2009molecular} \\ \hline
Relaxed Cell Perimeter $P_0$ & $3.6\sqrt{A_0} \mu\text{m}$ & Assumes circular resting shape \\ \hline
Cell-cell Contact Energy $\alpha$ & 1.0 & Sets energy scale \\
Cell-ECM Contact Energy $\beta$ & 3.5 & $2\beta > \alpha$ for cell adhesion \cite{graner1992simulation} (Varied in Fig.\ \ref{fig:heat}) \\ \hline
Area Energy Cost $\lambda_A$ & 1.5  & Prevents ``stringy'' cell-shapes \\
Area Energy Cost $\lambda_P$ & 0.01 & \\ \hline
\end{tabular}
\caption{Table of parameter values. Energy costs are in units of $k_B T$, where $k_B T$ is the thermal energy of the CPM Monte Carlo scheme.}
\label{table:param}
\end{table*}

\subsection{Connecting Gradient Sensing to Cell Motility}

To describe collective migration, we integrate the output of multicellular LEGI gradient sensing with cell motility. Cells in motion have a distinct front and are polarized along the direction of the front to back. Cells within the cluster have their polarization biased by a combination of the LEGI readout and intercellular repulsion due to contact inhibition of locomotion (CIL). CIL is the phenomenon where cells that come into contact cease to form protrusions in the direction of contact \cite{mayor2010keeping}. This is a very simple way for cells to translate the noisy, error-prone gradient measurements into collective cell motility \cite{camley2016emergent,malet2015collective,theveneau2010collective}.

In order to connect sensing to motility, we couple individual cell polarization $\vec{p}$ to both the LEGI downstream readout $R$ and what we will call the cell's repulsion vector $\vec{q}$. The cell's polarization vector represents the desired direction of motion \cite{jilkine2011comparison} and modeling collective behavior using cell polarization has been done previously\cite{szabo2010collective,camley2016emergent}. Information about the cell's surroundings are naturally expressed by the repulsion vector $\vec{q}$ \cite{camley2016emergent}. The repulsion vector is representative of contact inhibition of locomotion (CIL) \cite{mayor2010keeping}. CIL demonstrates that cells are aware of their immediate surroundings. The repulsion vector for cell $k$ is a unit vector that points away from all of cell $k$'s neighbors.
\begin{equation}
    \vec{q}_k = \left( \frac{1}{\sum_{\langle j,k \rangle} L_{j,k}|\vec{x}_k - \vec{x}_j|} \right)
    \sum_{\langle j,k \rangle} L_{j,k} \left( \vec{x}_k - \vec{x}_j \right),
\end{equation}
where $L_{j,k}$ is the contact length made between cell $k$ and its neighboring cell $j$. In our model cell polarization will change as a function of time depending on a combination of the repulsion vector and the LEGI downstream readout,
\begin{equation} \label{eq:polarVec}
    \frac{d\vec{p}_k}{dt} = r \left[ -\vec{p}_k + \epsilon \frac{R_k}{\sigma_R} \vec{q}_k \right].
\end{equation}
The first term in Eq.\ \ref{eq:polarVec} models the decay of cell polarization. In the absence of any stimulus an individual cell will undergo a persistent random walk with a timescale $1/r$ \cite{szabo2010collective}. The second term acts to align or anti-align the cell’s polarization vector with the repulsion vector, with alignment strength $\epsilon$ based on the cell's readout $R_k$. The magnitude of $R_k$ is normalized by its standard deviation $\sigma_R$. The net effect is illustrated in Fig.\ \ref{fig:model}A.

In the presence of a gradient, cells on the edge near the lower-end of the chemical concentration will tend to be polarized into the cluster (Cell 1 in Fig.\ \ref{fig:model}A), whereas cells on the higher concentration edge tend to be polarized outwards (Cell 3 in Fig.\ \ref{fig:model}A). Cells in the center of the cluster (Cell 2 in Fig.\ \ref{fig:model}A) are on average unpolarized. The net effect is that the cells on the edges of the cluster will drive motion in the direction of increasing chemical concentration. It is important to note that in this model single cells are unable to chemotax since the multicellular LEGI mechanism requires more than one cell to detect a gradient, and similarly without neighboring cells there is no repulsion vector to bias the cell's polarization.

\subsection{Computational Implementation}

Computational simulations are conducted in order to understand the dynamics that evolve from the model of collective sensing and migration. The source code for the simulations can be found here \cite{varennes2016zenodo54980}. The implementation chosen is the Cellular Potts Model (CPM) \cite{graner1992simulation,swat2012multi} although other cellular automata models are possible as well \cite{ermentrout1993cellular,maire2015molecular,mente2015analysis}. The CPM is widely used for simulating cell-centric systems. Despite its relative simplicity, this computational implementation can qualitatively reproduce diverse biological phenomena \cite{maree2007cellular}. The CPM is a very good implementation for simulating systems wherein cell geometry is crucial to the dynamics of the system. Using CPM many studies, some involving cell polarization and mechanical-based coupling, successfully reproduce epithelial cell streaming, cell sorting, chemotaxis and collective migration \cite{maclaren2015models,kabla2012collective,szabo2010collective}.

In the CPM cells exist on a discrete lattice and are represented as groupings of lattice points. Simply-connected groups of lattice sites $x$ with the same integer values for their \textit{lattice label} $\sigma(x)>0$ comprise a single cell. The extracellular matrix (ECM) is labeled with the lattice label $\sigma(x)=0$. Cells have a desired size and perimeter from which they can fluctuate, and cells adhere to their neighboring environment with an associated adhesion energy. The energy of the whole system is the sum of contributions from adhesion $J_{i,j}$, area-restriction $\lambda_A$, and perimeter-restriction $\lambda_P$ terms,
\begin{equation}
    u = \sum_{\langle x,x' \rangle} J_{\sigma(x),\sigma(x')} + \sum_{i=1}^N \left( \lambda_A (\delta A_i)^2 + \lambda_P (\delta P_i)^2 \right),
\end{equation}
\begin{equation}
    J_{\sigma(x),\sigma(x')} =
    \begin{cases}
        0 &\sigma(x)=\sigma(x') \ \text{(within the same cell)}, \\
        \alpha &\sigma(x)\sigma(x')>0 \ \text{(cell-cell contact)}, \\
        \beta &\sigma(x)\sigma(x')=0 \ \text{(cell-ECM contact)}.
    \end{cases}
\end{equation}
The parameters $\alpha$ and $\beta$ characterize intercellular adhesiveness, and in order to ensure that it is energetically favorable for cells to remain in contact, we restrict $\beta > 2\alpha$ \cite{szabo2010collective}. $\beta$ represents the cell-ECM contact energy, a larger value corresponds to an ECM that is more difficult to traverse. Heterogeneities in the microenvironment could be represented by a spatially dependent $\beta$; here we take $\beta$ to be a constant. The area- and perimeter-restriction energy terms prevent cells from growing or shrinking to unphysical sizes as well as branching or stretching into unphysical shapes. Cells fluctuate in shape and size around the desired area $A_0$ and perimeter $P_0$ with $\delta A_i \equiv A_i-A_0$ (and similarly for $\delta P_i$). The resulting dynamics evolve from the minimization of the system's energy under thermal fluctuations.

Cell dynamics are a consequence of minimizing the energy of the whole system. This is a random process that is sensitive to thermal fluctuations and is modeled using a Monte Carlo process. In a system of $n$ lattice sites, one \textit{Monte Carlo} time step (MC step) is composed of $n$ \textit{elementary} steps. Each elementary step consists of an attempt to copy the lattice label of a randomly chosen lattice site onto that of a randomly chosen neighboring site as illustrated by the pink lattice site in Fig.\ \ref{fig:model}B. The new configuration resulting from the copy is accepted with probability $P$, which depends on the change in the system's energy accrued in copying over the lattice label,
\begin{equation} \label{eq:prob}
    P =
    \begin{cases}
        e^{-\left( \Delta u - w \right)} &\ \Delta u - w > 0 , \\
        1 &\ \Delta u - w \leq 0 .
    \end{cases}
\end{equation}
The term $\Delta u$ is the change in energy of the system due to the proposed lattice label copy. $w$ is the \textit{bias} term which acts to bias cell motion in the direction of polarization. The bias term in the CPM model is required in order for cell clusters to exhibit directed motion \cite{szabo2010collective},
\begin{equation} \label{eq:w}
    w = \sum_{k=\sigma(a),\sigma(b)} \frac{\Delta\vec{x}_{k(a \to b)} \cdot \vec{p}_k}{ |\Delta\vec{x}_{k(a \to b)}| |\Delta\vec{x}_{k(\Delta t)}|}.
\end{equation}
The summation in Eq.\ \ref{eq:w} is over the cells involved in the elementary time step: $a$ is the lattice site being copied, and $b$ is the lattice site being changed. The change in the cell's center of mass position during the elementary time step is $\Delta\vec{x}_{k(a \to b)}$, whereas $\Delta\vec{x}_{k(\Delta t)}$ is the cell's change in the center of mass during a MC step. The cell polarization vector $\vec{p}_k$ is updated at every MC step in accordance with Eq.\ \ref{eq:polarVec}. The dot product acts to bias cell motion since movement that is parallel to the polarization vector will result in a more positive $w$ which in turn results in a higher acceptance probability (Eq.\ \ref{eq:prob}).

In addition to calculating the energy of the system, at each MC step the $X$ and $Y$ molecule populations in each cell are sampled by solving Eq.\ \ref{eq:xss} and \ref{eq:yss}. In doing so our model accounts for fluctuations in molecule numbers, cell shape, and cell-cell contact. With this computational implementation cells on the edges of the cluster are polarized in the direction of increasing chemical concentration, and cells near the center of the cluster have no net polarization, resulting in collective migration in the direction of increasing chemical concentration.

\section{Results}

\begin{figure*}[t]
    \centering
        \includegraphics[width=0.7\textwidth]{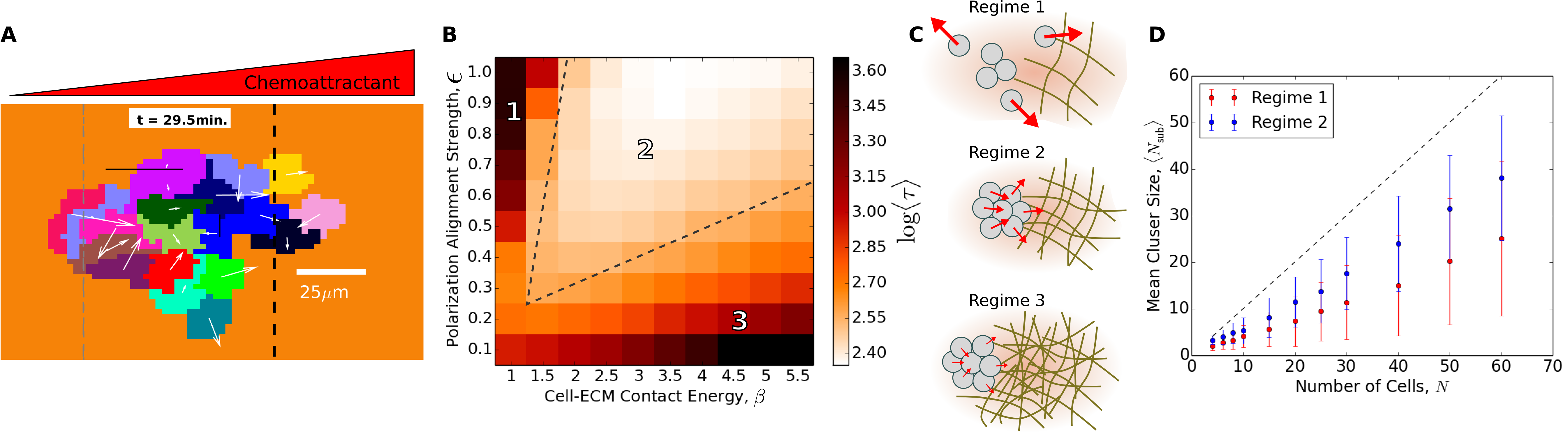}
    \caption{Characterizing the emergent multicellular migration. (A) Snapshot from simulation. Individual cells are distinguished by color and white arrows represent their polarization vectors. The cluster centroid is initially located along the gray dashed line and must cross the black dashed line in order to record a first-passage time event. (B) A heat-map of MFPT in units of minutes as a function of cell-ECM adhesion energy, $\beta$ and polarization bias strength, $\epsilon$. Warmer colors represent higher MFPT values (colorbar). Parameter values for the heat-map: $N = 20$, $\bar{c} = 10\text{nM}$, $g = 0.004\text{nM}/\mu\text{m}$, $\Gamma = 80 (\mu\text{m min.})^{-1}$. Illustrations in (C) represent cluster migratory behavior in their respective regimes of parameter space. Larger values of $\epsilon$ correspond to larger cell polarization vectors (red arrows), whereas larger values of $\beta$ correspond to an ECM that is more difficult to traverse. (D) Mean cluster size
    $\langle N_\text{sub} \rangle$
    as a function of the total number of cells in the system $N$. Regime 1: $\beta = 1.5$, $\epsilon = 1.0$. Regime 2: $\beta = 3.5$, $\epsilon = 0.8$} \label{fig:heat}
\end{figure*}

We simulate clusters of various sizes migrating in response to shallow constant chemical gradients over a fixed distance (Fig.\ \ref{fig:heat}A, Movie S1). The simulation results were calibrated using the cluster migration data from Malet-Engra et al.\ \cite{malet2015collective} and assuming a typical cell radius $a = 10 \mu\text{m}$. Similar to the experimental study, initial simulations were conducted with a gradient and background concentration equivalent to $\bar{g} = 0.001 \text{nM}/\mu\text{m}$ and $\bar{c} = 1 \text{nM}$. We found that increasing the gradient and background concentration values to those reported in Table \ref{table:param} (see pg.\ 11), which still maintain the limit
$a\bar{g} \ll \bar{c}$,
decreased computation cost while yielding the same qualitative results. Therefore all results presented here use the values of $\bar{c}$ and $\bar{g}$ in Table \ref{table:param}. The simulation timescale was then calibrated such that clusters of cells migrate with velocities on the same order as those in the study by Malet-Engra et al. All simulation parameter values used are presented and motivated in Table \ref{table:param} unless specified otherwise.

In order to quantify model behavior, statistics on the simulated mean first-passage time (MFPT) for migrating clusters are collected. The first-passage time is the time it takes for the center of mass of a cluster of cells to cross a threshold distance. First it is important to understand the effects of the various parameters in our model on simulations results. Across simulations, two crucial parameters emerge: $\beta$ the cell-ECM adhesion energy, and $\epsilon$ the polarization bias strength. When these two parameters are varied three distinct phases of collective cell migration are clear (regimes 1, 2, and 3 in Fig.\ \ref{fig:heat}B).

Fig.\ \ref{fig:heat}B shows that for sufficiently large $\beta$ the mean first-passage time remains relatively constant as $\beta$ and $\epsilon$ grow in proportion to one another. In this phase, regime 2 of Fig.\ \ref{fig:heat}B, cells migrate as a collective as illustrated in Fig.\ \ref{fig:heat}C. However if the adhesion energy is further increased while the bias strength remains fixed the MFPT starts to increase (regime 3 of Fig.\ \ref{fig:heat}B). This is due to the increased energy cost in cells making protrusions into the ECM. If $\beta$ is increased further the cluster cells will eventually stop moving since protrusions become highly improbable as dictated by the CPM (Fig.\ \ref{fig:heat}C). The other large MFPT phase is due to increasing $\epsilon$
while keeping $\beta$ fixed (regime 1 of Fig.\ \ref{fig:heat}B).
In this case the cell's polarization becomes large enough to overcome the intercell adhesion energy causing the cluster of cells to scatter as illustrated in Fig.\ \ref{fig:heat}C. To further characterize whether a cluster will scatter or remain persistently connected, we track the mean subcluster size $\langle N_\text{sub} \rangle$, defined as the average cluster size weighted by the number of cells present in each constituent cluster (Fig.\ \ref{fig:heat}D). Although cells' initial configuration is that of a single cluster, partial scattering may occur stochastically and reversibly, leading to a value of $\langle N_\text{sub} \rangle$ that is less than the cluster size $N$. As seen in Fig.\ \ref{fig:heat}D, the persistence $\langle N_\text{sub} \rangle/N$ is largely independent of $N$, and clusters in the parameter space of regime 2 are more persistent than those corresponding to regime 1 where cells are likely to scatter permanently. Overall, we see that there is a large region in parameter space which yields physically realistic behavior, and the model breaks down in the limits where we would expect it to. With this in mind we further examine simulations within regime 2 of parameter space.

\begin{figure}[t]
    \centering
        \includegraphics[width=0.45\textwidth]{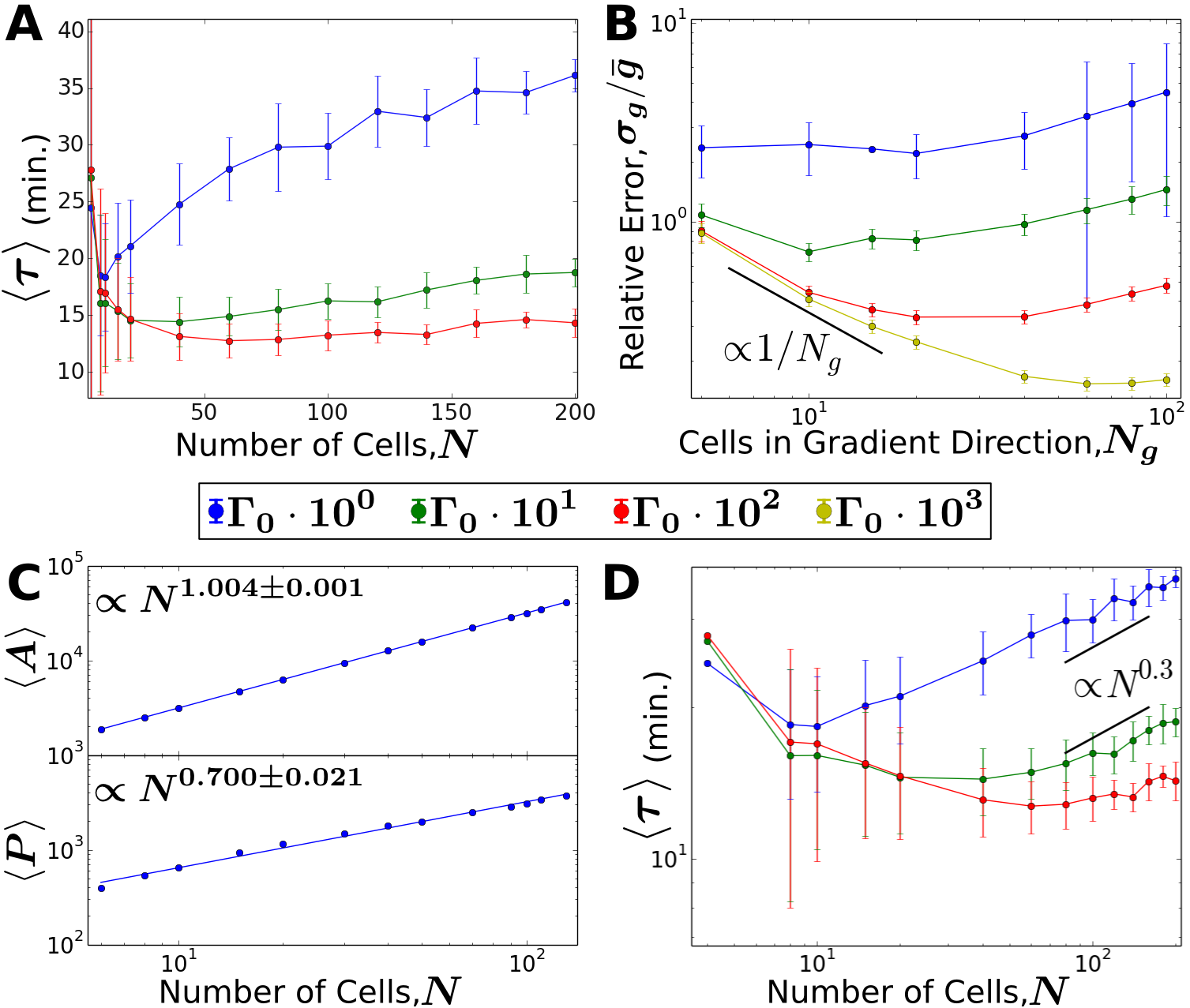}
    \caption{Tradeoff between sensing and drag leads to a minimum mean first-passage time (MFPT) with cluster size. $\Gamma_0 = 0.80 (\mu\text{m} \ \text{min})^{-1}$. (A) MFPT for various values of the exchange rate per unit contact-length $\Gamma$. (B) Relative error in gradient sensing for various values of $\Gamma$. (C) Area $A$ and perimeter $P$ scaling relationships with the number of cells $N$ in a cluster. (D) MFPT results in A on a log-log scale, compared with the geometric prediction arising from C. All error bars represent standard deviation.} \label{fig:results}
\end{figure}

Next we examine the MFPT as a function of cluster size (Fig.\ \ref{fig:results}A). Starting from $N=2$ we see that for sufficiently large $\Gamma$ (red curve), as the number of cells increases the MFPT decreases. This can be understood from our description of multicellular sensing (Eq.\ \ref{eq:G1}): before reaching the critical number of cells in a cluster, the error in gradient sensing decreases as $\sigma_R/\bar{R} \sim N^{-1}$ and so the cluster's ability to more precisely measure the gradient increases. The decreased sensing error translates into more accurately directed cell polarization vectors causing the MFPT to decrease. Fig.\ \ref{fig:results}B shows the relative error vs.\ the number of cells in the cluster that are parallel to the gradient direction, $N_g$. In the small-cluster regime and for fast communication (yellow curve) there is a decrease in relative error with $N_g$, that is in close agreement with the theoretical prediction for the scaling of $N_g^{-1}$
(Eq.\ \ref{eq:G1}). Since the global-reporter exchange rate between cells is very large compared to the degradation rate
($\gamma\gg\mu$)
it is expected that the effects of communication can be neglected as was the case in deriving Eq.\ \ref{eq:G1}. However, as the cluster grows in size the effects of communication can no longer be neglected. As illustrated in Fig.\ \ref{fig:results}B the relative error reaches a lower limit as predicted by Eq.\ \ref{eq:G2} at which sensory precision will no longer increase with increased cluster size.

As the number of cells increases the MFPT tends to saturate to a minimal value and may even begin to increase (Fig.\ \ref{fig:results}A). The MFPT reaches a minimum around $N \sim 10-100$ cells depending on the choice of $\Gamma$, the global molecule exchange rate per unit contact-length. Communication between cells improves as $\Gamma$ increases since more $Y$ molecules can be quickly transmitted between cells, pushing the point of saturation to larger cluster sizes. From these results we see that the model predicts an optimal cluster size for fastest migration. This prediction is in contrast with similar studies which in some cases predict a saturation in velocity and therefore constant MFPT as a function of cluster size \cite{camley2016emergent,malet2015collective}. The dependence of MFPT on cluster size is further explored in the Discussion.

In the limit that $\Gamma a/\mu \lesssim 1$ ($a$ being the cell radius) intercellular communication within the cluster is highly localized, and increasing the size of the cluster will not improve sensory precision. If this is the case then the cluster will have outgrown its optimal size for gradient detection. Instead of the cluster acting as one cohesive gradient-sensing device the cluster will comprise several independent gradient sensors which cannot reliably share information with one another. Therefore, in the small $\Gamma$ limit we expect the MFPT to monotonically increase with increasing $N$ due to increased drag on the cluster. Indeed, simulation results confirm our expectations in the large $N$, small $\Gamma$ limit (Fig.\ \ref{fig:results}A, blue curve).

Next we asked if the MFPT had any dependence on the geometrical properties of the migrating clusters \cite{camley2015collective}. The mean first-passage time should scale proportionally with the drag experienced on the cluster, whereas it should be inversely related to the force driving migration,
\begin{equation} \label{eq:fpt1}
    \langle\tau\rangle \sim \frac{\text{drag}}{\text{force}}.
\end{equation}
The drag on the cluster should scale with the area of the cluster, $\text{drag} \propto A(N)$, and the driving force should scale with the perimeter of the cluster since we know that only cells on the edges of the cluster are polarized in the desired direction, $\text{force} \propto P(N)$. Although the size and shape of clusters will fluctuate we can obtain from many simulations how the average area $\langle A \rangle$ and perimeter $\langle P \rangle$ scale with $N$.
Fig.\ \ref{fig:results}C shows that both scale with powers of $N$, i.e.\ $\langle A \rangle \sim N^d$ and $\langle P \rangle \sim N^f$. We find $d = 1.004 \pm 0.001$, which makes sense since the average area of the should scale linearly with the number of cells. We also find $f = 0.700 \pm 0.021$, which is intriguing because for a circular cluster we would expect $f = 1/2$. The larger value of $f$ reflects the elongated and amoebic shape of the cluster (Fig.\ \ref{fig:heat}A), which causes its perimeter-to-area ratio to be larger than that expected for a circle.

Given these geometric scalings, Eq.\ \ref{eq:fpt1} then makes a prediction: the MFPT should scale as $\langle\tau\rangle \sim N^{d-f} = N^{0.304 \pm 0.021}$. We compare this prediction to the MFPT data, on a log-log scale, in Fig.\ \ref{fig:results}D. We see that in the large $N$, small $\Gamma$ limit, the prediction agrees well with the data (blue and green curves). This demonstrates that the slowdown of large, poorly communicating clusters is dominated by the geometrical aspects of cluster propulsion and drag.

In summary, in the limit that communication between cells is strong ($\Gamma a / \mu \gg 1$), information can be reliably transferred over $n_0 \gg 1$ cells. As long as cluster sizes $N$ remain smaller than $n_0$ cells, there will be an improvement in the sensory capability of the cluster with size, and an associated decrease in the MFPT $\langle\tau\rangle$.  As the critical size $n_0$ is reached, sensory ability will cease to improve with size, and $\langle\tau\rangle$ will reach a minimum. Further addition of cells will cause $\langle\tau\rangle$ to increase according to
$\langle\tau\rangle \sim \text{drag}/\text{force}$, since the drag is proportional to the cluster area, whereas the force is proportional only to the cluster perimeter.

\section{Discussion}

We have developed a model in which collective sensing of noisy chemical gradients induces multicellular migration. The model includes the stochastic processes of ligand diffusion, intercellular communication and cell shape fluctuations.
In the model cells are polarized based on collective gradient information and contact-mediated interactions, leading to biased migration despite the fact that individual cells do not chemotax. We find that the antagonistic effects of sensing and drag result in a minimum mean first-passage time (MFPT) as a function of cluster size, i.e.\ an optimal size for fastest migration. The optimal size is governed by the strength of cell-cell communication, with stronger communication leading to both a larger optimal size and a decreased migration time (Fig.\ \ref{fig:results}D).

Whereas previous models have idealized cell or cluster geometries as perfect circles \cite{ camley2016emergent, camley2015collective}, our use of the cellular Potts model has allowed us to capture natural fluctuations in cell and cluster shape. As a result, we have found that while migrating, clusters adopt a shape that is (i) elongated in the gradient direction and (ii) non-convex (see Fig.\ \ref{fig:heat}A). Both features lead to a cluster perimeter-to-area ratio that is significantly larger than that expected for a circle or other convex shape with aspect ratio near unity. Importantly, we have found that the area and perimeter scalings remain predictive of MFPT in the communication-limited regime (Fig.\ \ref{fig:results}D), even with the observed non-circular and fluctuating geometries.

To the extent possible, our model has been constructed and parameterized using current experiments on collective migration. Intercellular communication is modeled as a direct exchange of messenger molecules between cells since this type of  communication has been implicated in development, organoid branching, angiogenesis, and cancer \cite{ramel2013rab11,ellison2016cell,gerhardt2003vegf,hsu2000cadherin,friedl2009collective}. The chemical concentration and gradient values are selected to ensure that our simulations are in the shallow gradient regime, where experiments show that collectives can respond whereas single cells cannot \cite{ellison2016cell,malet2015collective,rosoff2004new}. Cell size, chemical concentration, chemical gradient, cell-cell contact energy, and cell-ECM contact energy values are taken from previous experimental studies of collective cell behavior (Table \ref{table:param}).

How do our model predictions compare to experiments? There have been many studies on collective migration \cite{cheung2013collective,puliafito2015three,defranco2008migrating,ramel2013rab11} though only one (to our knowledge), by Malet-Engra et al.\ \cite{malet2015collective}, measures migratory properties as a function of cluster size. The experiments conducted by Malet-Engra et al.\ reveal that beyond a minimum cluster size, the cluster velocity saturates to a maximal value and then remains constant with increasing cluster size. In our study, we find that when communication is strong, the MFPT -- which is inversely related to the mean velocity -- also saturates to a minimal value and remains constant for a large range of cluster sizes. As shown in Fig.\ \ref{fig:results}A (red curve), as the cluster size increases from about 30 to 200 cells the MFPT remains relatively constant, in qualitative agreement with the aforementioned experimental results. This saturation regime occurs when communication is sufficiently strong to suppress, over a large range of cluster sizes, the drag-induced slowdown. Our findings thus suggest that sensory information is reliably transferred throughout the clusters of lymphocytes studied by Malet-Engra et al., and that communication is strong enough that drag does not strongly constrain migration speed for the cluster sizes analyzed.

\begin{figure}[t]
    \centering
        \includegraphics[width=0.45\textwidth]{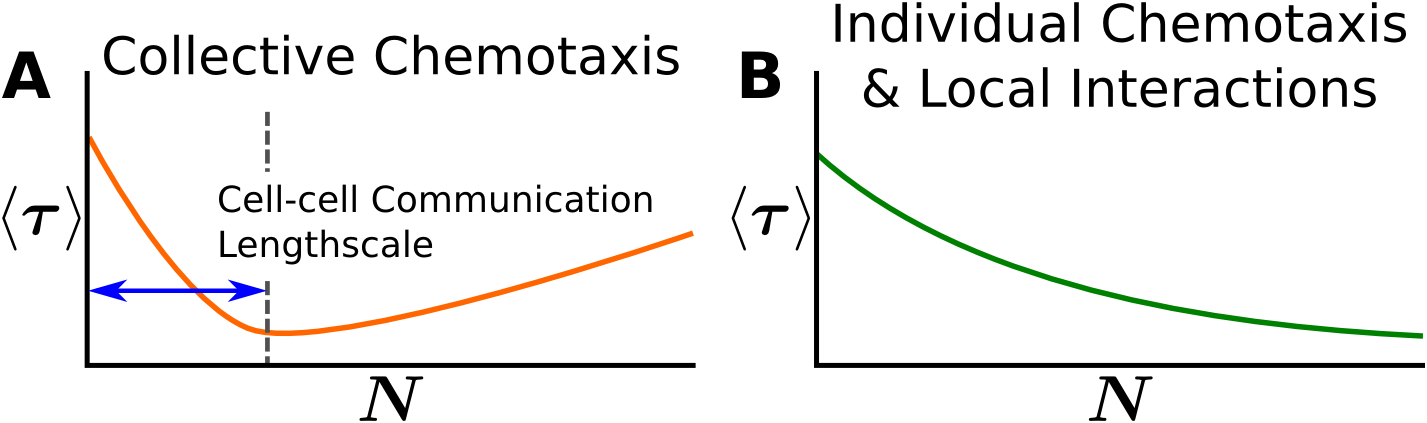}
    \caption{Prediction to distinguish collective from individual chemotaxis in experiments. (A) Expected MFPT behavior for cluster migration driven by collective sensing. (B) Expected MFPT behavior for cluster migration driven by local interactions.} \label{fig:mfpt}
\end{figure}

Furthermore, our results suggest a simple experimental test that can distinguish whether cluster chemotaxis is purely collective or individually driven. Broadly speaking, cluster migration (i) can emerge collectively from cells that communicate, either chemically or mechanically, but do not chemotax alone (as in our model), or (ii) it can result from many individual agents that take independent measurements of the environment and through physical coupling or local interactions produce collective migration \cite{coburn2013tactile,vicsek1995novel} (a so-called ``many wrongs'' mechanism \cite{simons2004many}). As illustrated in Fig.\ \ref{fig:mfpt}A, our results suggest that in the former case, one would observe a minimum in the migration time as a function of the cluster size, with the optimal size determined by the length scale of collective information processing within the cluster. In contrast, as illustrated in Fig.\ \ref{fig:mfpt}B, in the latter case migration is driven by the integrated measurements of many effectively independent agents, and thus one would observe a monotonic decrease in the migration time as a function of the cluster size \cite{simons2004many}. Distinguishing the dependence in Fig.\ \ref{fig:mfpt}A from that in Fig.\ \ref{fig:mfpt}B using microscopy would provide phenomenological evidence of purely collective chemotaxis without relying on molecular-level details.

An important feature of our model and its analysis is that the timescale of sensing is faster than the timescale of cell response and motility (Table \ref{table:param}). However, in actuality the duration of cells' sensing timescales relative to their response timescales is unknown \cite{ellison2016cell}. If the motility timescale is shorter than that of sensing for a specific cell type than the MFPT dependence on cluster size may be more complicated than predicted. For short response timescales we expect migratory behavior to be more strongly diffusive, but to still remain biased in the direction of the gradient over periods of time larger than the sensing timescale.

In our model, the precision of multicellular migration is determined in part by noise arising from ligand diffusion at the initial sensory stage. As such, the model respects the fundamental limits to the precision of collective gradient sensing set by the physics of diffusion, which were recently tested in collectives of epithelial cells \cite{ellison2016cell, mugler2016limits}. It will be interesting to see how these and similar limits translate from the domain of sensing to that of migration, and whether they depend on the underlying migration mechanism (purely collective, individually driven, or a mixture thereof).

\section*{Author Contributions}
J.V.\, B.H.\ and A.W. designed the research and analyzed the data; J.V.\ performed the research; J.V.\ and A.W.\ wrote the manuscript.

\section*{Acknowledgements}
This work was supported by the Ralph W. and Grace M. Showalter Research Trust. J.V.\ was additionally supported by the Purdue Research Foundation.

\end{document}